\documentclass[pt12]{article}
\begin{document}
\begin{center}
                                           Quantisation on general spaces\\

                                                   Ajay  Patwardhan \\

               Physics Department , St Xavier's college ,Mumbai 400001, India\\

                                        quant-ph, math-ph\\

                                            8th   November 2002\\
\end{center}

 Quantisation on spaces with properties of curvature ,
 multiple connectedness and non orientablility is obtained. The geodesic
 length spectrum for the Laplacian operator is extended to solve the
 Schroedinger operator. Homotopy fundamental group representations
 are used to obtain a direct sum of Hilbert spaces, with a Holonomy
 method for the non simply connected manifolds.The covering spaces
 of isometric and hence isospectral manifolds are used to obtain
 the representation of states on orientable and non orientable
 spaces. Problems of deformations of the operators and the
 domains are discussed.Possible applications of the geometric
 and topological effects in  physics are mentioned.

\section{}
{
 
                                INTRODUCTION

The Mathematical Quantum Physics on spaces with curvature, multiple connectedness and non orientability is an important topic for many subjects. Schroedinger free particle quantisation on such spaces involves, solving the Helmholtz equation. The problem with interactions can be developed with Lipmann Schwinger methods once the basis Hilbert space is defined.

 The problem of solving differential equations on domains with shape deformations, for example, from a sphere to a spheroid to an oval shaped space with some holes and handles, is a classical one. Progress towards  solutions of these kind of problems have required modern mathematical methods of spectrum of operators on Riemann manifolds, with geometric and topological properties.

A geodesic length spectrum of the operator $( \nabla^{2} +k^2 )$ is to be used.Kernels are obtained to define the quantum states. For manifolds with genus, the Hilbert space becomes a direct sum of Hilbert spaces, given by the fundamental group representation on the manifold.  Covering spaces are used for both the orientable and non orientable cases. The framework is in principle applicable to spaces with all the three properties. Using isospectral manifolds with genus, it can be extended to orientable and non orientable spaces. The annular disc and the Mobius strip are the prototypes.

The applications of the operator spectrum and properties based on it, could be of interest in solid state thin films and Fermi surfaces,in molecular systems and cavity resonators; and to spaces occuring in string, field and gravity problems. They can be related to methods used of path integrals and partition functions and holonomy in these subjects.In spaces with symmetry the methods of special functions and Green functions are usually used to solve the spectrum of differential operators.

 The methods in this paper have been developed in the past twenty years in a variety of sub-fields of mathematics and physics, so that solving for general spaces can now be done [References 1 to 14].

}
\section{}

{  ISOSPECTRAL  MANIFOLDS : THE SPECTRUM OF OPERATORS;GEODESIC LENGTHS ,HOMOTOPY AND HOLONOMY.

Isometric manifolds are isospectral, but the converse is not always true. Two manifolds are isospectral if and only if their Zeta functions are equal.\qquad $ Z_{M1}(s) =  Z_{M2} (s)$

 If $ l_p $ is the length of the pth geodesic cycle (equivalence class of geodesics) then the Zeta function is defined as

 $ Z_M (s)  = \prod_p [1 - [exp(l_p ) ]^{-s} ]^{-1} .$

$ Tr  exp[-t(\nabla^2   +  k^2 ) ] =  \sum_{\lambda_n} exp(-t\lambda_n)$\qquad is the trace formula.

The Laplacian operator on a Riemann manifold with metric h is given by, in coordinates $ x^i$:

 $\frac{1}{\sqrt{h}}  \frac{\partial}{\partial{x^i}} (h^{ij} \sqrt{h} \frac{\partial}{\partial{x^j}}) $ where h stands for $ det(h_{ij})$

$K_M (x,x')  =  \sum_{n=0}^\infty   K_n  \psi_n (x)  \psi_n (x')$    \qquad is the kernel.

$K_n  =  \frac{1}{k_n^{2} -k^{2} }$    \qquad     $SP(K_M)  =  \sum_{n=0}^\infty K_n $\qquad is the spectrum.

For a Laplacian on a finite collection of isospectral manifolds $(M) $ , $L(M) $ is the length spectrum of equivalence class of closed geodesics

  $SP(M) =( +/_-) 2 Cosh (\frac{1}{2} L(M))$ is the spectrum.

$ The shortest closed path * diameter(M) = area (M) = 4\pi(g - 1)$   for the Riemannian space M of genus g, where  genus is the \# holes - \#handles.  For multiply connected spaces, super selection sectors of the direct sum of Hilbert spaces arises with the fundamental group representation index.

$K(\phi,t;\phi_0,t_0)  = \sum_{n=-\infty}^\infty\sum_\phi[exp(iS\phi_n (t))] $ is a path integral.

$K = \sum_n  A_n K_n $   \qquad   $ A_{n+1} = exp(i\delta) A_n  $  with $  A_n =exp(in\delta)$  where $ \delta$ is a real phase.

$K_n = \left(\frac{I}{2\pi{iT}}\right)^\frac{1}{2} exp[\frac{iI}{2T}(\phi - 2n\pi)^2]$

is the kernel for a single exclusion with   $ L =\frac{1}{2} I \phi'^{2}$ and $ T = t-t_0.$

This gives a winding number phase and $ E_J = \frac{1}{2I} (J +\frac{\delta}{2\pi})^2 $ gives the quantisation rule.

For Abelian circle group and cyclic groups a phase  $exp [i\sum_{k} \theta_k n_k] $ is taken . Introducing a one form$ A_k$

 and $\frac{2\pi e_k}{hc}\oint A_kdx =  n_k\theta_k $ over the  distinct loops ; a representation for the Abelian Fundamental group is generated.

More generally a Holonomy is defined for non Abelian fundamental groups $\pi(M)$ on $ A_k(M)$ forms. With this the spectral conditions could be extended to nontrivial general spaces. The use of the forms A is not an introduction of gauge fields ,but as realisations of the homotopy groups to enable the classification of allowed inequivalent states in Hilbert space. The subject of gauge field equations on manifolds has developed holonomy methods, which are adapted to this problem.

The general formula is written as \qquad$Tr P exp[\sum_{k} C_k \oint_M A_kdx]$

The homotopically equivalent paths of closed loops on the manifold M with genus g are used in the integral, giving rise to a Fundamental group representation.

The Unitary irreducible representation of the path group  $\pi_1 (M)$ gives the equivalence class of basis states . If $ \pi_1 (M)$ is not the identity, then the manifold M is multiply connected.

 The representations for the $ \pi (M) $ groups are computable for many examples of manifolds in algebraic topology.

 In the example above the $\pi_1(M) = Z $.  Then the Hilbert space is the direct sum of Hilbert spaces with the Fundamental group representation index.

For non abelian fundamental groups a multiplicity of states occurs. The covering space of all equivalence classes of the same FG index( fundamental group index) is simply connected. The Homology group is also used to find nontrivial realisations, for example when the $ \pi_1(M)$ is the Braid group for more than two particles. One and higher dimensional Unitary irreducible representations have been found.

 For Abelian $ \pi_1(M)$ an irreducible representation occurs. On each subspace there is ``inequivalent quantisation''with different index values $ l$ and $ m $ , for the paths $ \gamma_x ^l (t)  <-->  H_L$ and $ \gamma_x ^m (t) <--> H_m.$, with $ x \varepsilon M $. 

 For isometric smooth deformations of the manifold, these Hilbert space sectors will remain invariant.

For topology changing transformations of the manifold, the Euler number  $ \chi(M) = 2-2g $ is an invariant, given by the sum of betti numbers of M  : $ \sum_k(-1)^k b_k(M) $. 

 This will control the superselection sectors, given by the FG index ,and give transition probabilities among the inequivalent quantisation sectors.In summary  the Total Hilbert space is given as

$H = \bigoplus_{FGindex_n}[ H_{FGindex_n}]$

}

\section{}
{THE COVERING SPACES FOR THE MULTIPLY CONNECTED MANIFOLD WITH CURVATURE AND ORIENTABLE AND NON ORIENTABLE SPACES.

Two manifolds $ M_1(h_1)$ and$ M_2(h_2)$ are isospectral for the  operator  $ \nabla^2 + k^2 $ , if and only if their Zeta functions :  $  \sum_{i=0} ^\infty \lambda_i^{-s}$  with  $ s>0$  and $  0<\lambda_1<\lambda_2 .......,$ ; are equal for the two manifolds. Isometric manifolds are isospectral. The number of isospectral manifolds increases rapidly with the genus number.

 The collection $ M_i$ of manifolds, with covering spaces $ C_i$ is such that $ C_i : M_i --> M_0 $ with $\bigcup_i{C_i}$ as the total covering space of the manifold M. A convenient manifold $ M_0$ which is isospectral to the original manifold M is chosen in this construction.

 The metrics $ h_i$ on $ M_i$ induce the metric $ h_0$ on $ M_0$ such that $ M(h)$ and $ M_0(h_0$) are isometric, and hence isospectral. The spectral problem can be solved on the convenient $ M_0$ and then lifted back to the manifold M.

When a transformation group G acts on M, with subgroups $H_1, H_2, H_i$ etc; the covering spaces $C_i$ of the sub manifolds $ M_i = \frac{M}{H_i}$ are induced. Finitely many covering spaces exist, $C_i$ of the manifolds $M_i$ such that $\bigcup_i{C_i}$ covers the manifold $ M $, with each $C_i$ as a simply connected space.

 Riemann surfaces exist for genus $ g\geq 2$. The analogy is to'' rubber patches'' covering distinct subspaces, which can then be mapped onto domains on which the operator spectrum is easier to compute.

For orientable manifolds the choice of covering spaces  is : ${C_i}\bigcap{C_j}$ is null, and for the non orientable spaces it is not null( for some i,j). A deRham integration measure can be used to take care of both cases.

 Projective spaces equivalent to the non orientable spaces are found and used. For example,the Fundamental group is  $ Z_{n+1}$ for a 'n' twist Mobius band, and gives a winding number index for the phase.

The prototypes of multiply connected and non orientable spaces are respectively the annular disc and the Mobius strip with a single twist. The decomposition of general spaces into topologically equivalent spaces to these two exists.

 The boundary conditions are set on the covering spaces $ C_1$ and $ C_2 $, with ${ C_1}\bigcup{ C_2}$ . For the annular disc choose : $ \psi_1(C_1) = \psi_2( C_2)  exp(2i\pi)$, with ${C_1}\bigcap{C_2} $ null; and for the Mobius strip set $\psi_1(C_1) = \psi_2(C_2)  exp(i\pi) $, with $ { C_1}\bigcap{C_2} $ not null.

To summarise the Quantisation framework on general spaces that are decomposable into a collection of orientable and nonorientable spaces. The total Hilbert space is a direct sum of Hilbert spaces on the decomposition of manifolds.

 So the general manifold M , with genus g and Euler number $\chi(M) = 2-2g $ as conserved ; has a covering space $\bigcup_i{C_i}$ with ${C_i}\bigcap{C_j} $ null for orientable, and non null for non orientable cases.

The `` convenient'' isospectral manifold for computation of the geodesic length spectrum is $M_0$ with the isometric condition with the original manifold M. This is ensured by the subgroups $H_i$ of a group G of general transformations of the manifold, such that $ M_i =\frac{M}{H_i}$ and $C_i$ cover $M_i$. The metrics $ h_i$ on $M_i$ induce the isometric $h_0$ on $M_0$ to the metric h on M.

The quantisation with the geodesic length spectrum on each $M_i$ then induces the quantisation on M.With the conveniently chosen $M_0$ , where exactly computable geodesic length spectrums are possible, either analytically or numerically; the spectral conditions on $M_i$ and $M_0$ can then be lifted back to the spectral condition for the original manifold M. The Zeta functions and the Kernels are calculated and the geodesic lengths give the allowed reciprocal wave vectors ``k'' and hence the energy spectrum for the free particle.

For specific, compact,connected , isospectral manifolds , with all eigenvalues of the( Laplacian + $k^2$) operator to be found; the problem is now reduced to essentially the problem of computing equivalence class of geodesic lengths with a convenient metric, as done in differential geometry.This paper has made a framework based on work that is reported in a number of distinct subjects.

}

\section{}
{                    VARIATIONS OF THE BASIC PROBLEM

 For deformations of the manifolds that break isometricity conditions, the sensitivity of the spectrum of the operator to these changes is a subject by itself. Known results for isospectral polyhedral domains and symmetric polytopes could be used to test this framework. Some isospectral deformations are known. Does the spectrum determine the geometry and the operator completely. Does the geometry determine the spectrum uniquely; these are active research questions with some answers known.The spectrum of the Laplacian on a manifold may not determine its topology and vice versa, as seen in index theorems.

 The domain and boundary dependence for symmetric manifold cases is usually seen from the appropriate  symmetry group representations with special functions as the basis. In terms of these functions, using their linear superpositions, the effect of small perturbations of the operators and their domains could be studied. This has the effect of lifting degeneracy or suppressing some states in the spectrum of the operators.There are Riemann manifolds so that Laplacian plus two potentials have same spectrum , and spaces can be found so that the spectrum of Laplacian plus a potential uniquely determines the potential.With this the Ricciscalar term $\frac {R}{4} $ can be added to the Riemann manifold Laplacian.  Other operators ,such as the Dirac operator with gauge potentials are also being studied on specific manifolds.

 For large deformations only numerical methods were useful. Two kinds of results are significant.
(1) The occurence of chaos and also integrable solutions : (a) Scarring of wavefunctions, (b) complex patterns of nodal subspaces, (c)changes of statistics of energy level spacings , from integrable to chaotic type. $ P(s) = exp(-s)$ changes to $P(s) = s^m exp(-s^n)$ with $ s=\frac{\Delta{E}}{<\Delta{E}>}$. These effects occur in quantum billiards, for example.

(2) Occurence of isospectral domains , spaces and manifolds. Sets of isospectral domains have been found , for example in microwave cavity studies, and in the polyhedral geometries. To obtain a general rule for finding them is an active subject of research ; for numerical, experimental and analytical work. For example, quantum optics requires cavities of various shapes and allowed modes.

The approach of this paper to the problem can have application to classical mathematical physics of accoustics, electrodynamics, heat conduction and wave propagation and waveguides in which these shape dependent effects would be required. Algebraic and numerical computing of Differential geometry and graphic display ( for example with Mathematica)  has opened a range of possibilities.

For solving deviations from the Helmholtz operator, the methods of integral equations can be used. The free particle Hamiltonian on the manifold is then replaced by a interacting Hamiltonian (including potential energy terms). Given the basis states and eigen values, the Lipmann Schwinger equation can be used to generate the full solution in a perturbative analysis.

The free particle spectrum of $(\psi_0, E_0)$ is used to obtain the new states.

 $ \psi ={ \frac{1}{I-GH'}} \psi_0 $ with $ G = (E-H_0)^{-1} $, where $ H = H_0 + H'$ and $ H_0 \psi_0 = E_0 \psi_0$, also $ H\psi = E\psi$. A order by order calculation can be done for bound states (perturbation) and unbound states(scattering).

 As a first step a convenient potential H' is taken. Then a succession of perturbations is formally solved by taking $ \psi_0 --> \psi$ and $ \psi --> \Psi'$, with $ H_0 --> H_0 + H' --> H_0 + H' + H'' $; and applying the Lipmann Schwinger formalism over again. 

Hence the construction of the problem with variations of the operator as well as the domain  has been  outlined. There will be questions such as selfadjointness and convergence to consider in specific examples. How sensitive is the spectrum to the variations and what physically interesting effects can arise is also a question.

}

\section{}
{                                        MORE   APPLICATIONS

In mesoscopic systems the shape, size and geometry dependent effects are observed as important. In thin films or ribbons with curvature ,multiple connectedness and non orientability, all the properties could be studied. In devices etched with shape variations and deformations like kinks, exclusions and bumps it is possible to create useful and novel effects .The classification of the electron energy levels is an important step towards understanding conduction problems.

Fermi surfaces (in ``k'' space) have curvature and topology (genus), where excluded ``k'' vectors are ``holes'', and ``k'' vectors identified as equivalent are ``handles' in ``k'' space. Transformations of the material can create non trivial and interesting changes in these properties of the Fermi surfaces. For example,' pinched' handles and 'blocked' holes' ; and non smooth Brillouin zone boundaries.

 Methods of analysis of the electron/hole spectrum on Fermi surfaces therefore gives a way to understand these effects for conduction problems in solid state and mesoscopic physics. Fullerene like molecules with shape deformations and molecules with nonorientable surfaces as they occur in protein folding are other likely applications.

 A thin film with local exclusions can give easy to test examples of multiply connected surfaces.In magnetic superconductors these naturally occur.In quantum systems, modifications of standard problems or a new class of problems could be identified. This subject is relatively at a beginning stage and much more work may be expected in the future.

In gravitational systems with horizons, in string and field theories on manifolds with geometric and topological properties, and in many particle systems on manifolds; there is growing work using path integrals, partition functions and Holonomies, with fundamental homotopy group representations. The introduction of noncommutative and quantum geometries takes the issues of quantising on spaces, with geometry and topology dependent effects to a fundamental level. The spectrum of operators on these general spaces is a topic of current research. 
}
\section{}
{ACKNOWLEDGEMENTS

The author gratefully acknowledges the hospitality and facilities at the following Institutions. The Center for Theoretical Studies at Indian Institute of Science, Bangalore (1996),and Central University at Hyderabad ( 1997) where this work was begun. The work was reported by me in the International association of Mathematical Physics conference at Department of Mathematics ,Nagpur University (1999) and in the Winter Institute on Foundations of Quantum Theory at the S.N.Bose National center for Basic Sciences, Kolkata (2000). The work is completed in this form at Institute of Mathematical Sciences ,Chennai (2002).

}

\section{}

                                      {    REFERENCES

(1) D. Webb,S.Wolpert, C. Gordon ; B.A.M.S. 27, 134-138,(1992)

(2) T. Sunada; Ann. of Math 121, 169-186,(1985) 

(3) Isaac Chavel,  Eigenvalues in Riemannian geometry; Acad Press(1984)

(4) T.Imbo, E.C.G. Sudarshan ,T.R.Govindarajan; Physics Letters 213B \qquad  (1988)

(5) A.P.Balachandran,G.Marmo,B.Skagerstam, A.Stern;Classical Topology \qquad  and Quantum states WSP(1991)

(6) G. Morandi ,Topology in classical  and quantum physics;(Springer Verlag)

(7) D.Webb , C.Gordon ,Am Sc Jan 96

(8) R.Brooks , R. Gornet , W.Gustafson ; Adv. in Math 138,306-322 (1998)

(9) P.Horvathy , G.Morandi, E.C.G.Sudarshan , Il Nuovo Cimento  110 ,201 \qquad  (1989) 

(10) T.Imbo , C.Imbo , E.C.G.S.Sudarshan , Phys Lett B 234,103 (1990) 

(11) P.Buser, Geometry and spectrum of complex Riemann surfaces, Birkhauser (1992)

(12)Giampiero Esposito , Dirac operators and spectral geometry ,Cambridge University Press(1998) 

(13) S.Rosenberg The Laplacian on a Riemannian manifold,London Math soc31, Cambridge university press(1997)

(14)R.Brooks , Am.Math.Monthly 95,823-839 (1988)

}

\end{document}